%% file: ISTC_FAID_NN.tex
\documentclass[conference]{IEEEtran}
\IEEEoverridecommandlockouts
\usepackage{cite}
\usepackage{amsmath,amssymb,amsfonts}
\usepackage{algorithmic}
\usepackage{graphicx}
\usepackage{textcomp}
\usepackage{xcolor}
\usepackage{epstopdf}

\usepackage{soul}
\setstcolor{red} %use command \st for striking out text
\setul{0pt}{0.7pt} %line weight
\newtheorem{proposition}{Proposition}
\def\beginofproof{\noindent{\it Proof: }}
\def\endofproof{\hfill\rule{6pt}{6pt}}
\def\BibTeX{{\rm B\kern-.05em{\sc i\kern-.025em b}\kern-.08em
    T\kern-.1667em\lower.7ex\hbox{E}\kern-.125emX}}
\begin{document}

%\title{Decoder Diversity of Finite Alphabet Iterative Decoders for LDPC codes via Neural Networks \\
\title{FAID Diversity via Neural Networks \\
%change the "Intelligent"
\thanks{The work is funded in part by the NSF under grants CIF-1855879, CCF 2106189, CCSS-2027844 and CCSS-2052751. Bane Vasi\'c has disclosed an outside interest in Codelucida to the University of Arizona.  Conflicts of interest resulting from this interest are being managed by The University of Arizona in accordance with its policies.}
}

\author{\IEEEauthorblockN{Xin Xiao\IEEEauthorrefmark{1}, Nithin Raveendran\IEEEauthorrefmark{1}, Bane Vasi\'{c}\IEEEauthorrefmark{1}, Shu Lin\IEEEauthorrefmark{2}, and Ravi Tandon\IEEEauthorrefmark{1}}\\
 \IEEEauthorblockA{\IEEEauthorrefmark{1} The University of Arizona, Tucson, AZ, 85721. Email: \{7xinxiao7, nithin, vasic, tandonr\}@email.arizona.edu}\\
 \IEEEauthorblockA{\IEEEauthorrefmark{2}University of California, Davis, CA, 95616. Email: shulin@ucdavis.edu}
}
% \author{\IEEEauthorblockN{Xin Xiao}
% \IEEEauthorblockA{\textit{Electrical and Computer Engineering} \\
% \textit{The University of Arizona}\\
% Tucson, United States \\
% 7xinxiao7@email.arizona.edu}
% \and
% \IEEEauthorblockN{Nithin Raveendran}
% \IEEEauthorblockA{\textit{Electrical and Computer Engineering} \\
% \textit{The University of Arizona}\\
% Tucson, United States  \\
% nithin@email.arizona.edu}
% \and
% \IEEEauthorblockN{Bane Vasi\'c}
% \IEEEauthorblockA{\textit{Electrical and Computer Engineering} \\
% \textit{The University of Arizona}\\
% Tucson, United States \\
% vasic@ece.arizona.edu}
% \and
% \IEEEauthorblockN{Shu Lin \red{\hlc{why is the second row not centered?}}}
% \IEEEauthorblockA{\textit{Electrical and Computer Engineering} \\
% \textit{University of California, Davis}\\
% Davis, United States \\
% shulin@ucdavis.edu}
% \and
% \IEEEauthorblockN{Ravi Tandon}
% \IEEEauthorblockA{\textit{Electrical and Computer Engineering} \\
% \textit{The University of Arizona}\\
% Tucson, United States \\
% tandonr@email.arizona.edu}
% }

\maketitle

\begin{abstract}
\emph{Decoder diversity} is a powerful error correction framework in which a collection of decoders collaboratively correct a set of error patterns otherwise uncorrectable by any individual decoder. In this paper, we propose a new approach to design the decoder diversity of finite alphabet iterative decoders (FAIDs) for \emph{Low-Density Parity Check} (LDPC) codes over the binary symmetric channel (BSC), for the purpose of lowering the error floor while guaranteeing the waterfall performance. The proposed decoder diversity is achieved by training a recurrent quantized neural network (RQNN) {to learn/design FAIDs}. {We demonstrated for the first time that a machine-learned decoder can surpass in performance a man-made decoder of the same complexity.} As RQNNs can model a broad class of FAIDs, they are capable of learning an arbitrary FAID.  
%\hlc{iteratively} 
%\st{which is a model-driven }
%\hlc{model-driven} 
%\red{[not clear, why do we need this concept of model-driven nn to explain what we do?]} 
%\st{neural network capable of learning any arbitrary FAID.}
%\hlc{capable of learning any arbitrary FAID} 
%\red{[not clear, why can learn arbitrary faid?]}. 
To provide sufficient knowledge of the {error floor to the RQNN,} %\st{trapping sets} \red{[what are trapping sets? they are not introduced and we tald about them]}, 
the training sets are constructed by sampling from the set of most problematic error patterns - trapping sets. %via the sub-graph expansion-contraction --  a semi-analytical,  computationally efficient method to estimate the error floor {-- so that the learned FAIDs are optimized in the error floor region.}
%\red{\hlc{How is error floor  related to learning?}}. 
In contrast to the existing methods that use the cross-entropy function as the loss function, we introduce a frame-error-rate (FER) based loss function to train the RQNN with the objective of correcting specific error patterns rather than reducing the bit error rate (BER). 
The examples and simulation results show that the RQNN-aided decoder diversity 
%not only performs well in the waterfall region, but also 
increases the error correction capability of LDPC codes and lowers the error floor. %significantly.%rather than randomly sampling at a high crossover probability.%The proposed decoder diversity is designed by training a recurrent quantized neural networks (RQNN) iteratively, whose training sets are constructed by sampling the most problematic error patterns via the sub-graph expansion-contraction. %a semi-analytical and computationally efficient method. 
%The RQNN is a model-driven neural network, which is capable of learning any arbitrary FAID, and the sub-graph expansion-contraction is a semi-analytical method to estimate the error floor efficiently.
\end{abstract}


%Long version
\input{Introduction}

\input{Preliminaries}
\input{Intellgient_Decoder_Diversity}
\input{Results}
\input{Conclusion}

%Short version
% \vspace{-8pt}
% \input{Introduction_short}
% \vspace{-8pt}
% \input{Preliminaries_short}
% \input{Short_Intellgient_Decoder_Diversity}
% \input{Results_Short}
% \vspace{-8pt}
% \input{Conclusion}

%TO-DO:

%Outline:

%1) all the existing model-driven NN optimize in waterfall region;

%2) Decoder diversity optimize both, and no single FAID can perform well in both waterfall and error floor;
%Section I:

%3) Sub-graph expansion and contraction;

%4) Main contribution of this paper.

%Section II:

%0) notation of Trapping sets;

%1) definition of general FAID;

%2) definition of decoder diversity

%3) introduction of RQNN for linear FAID.

%Section III:

%1) whole framework;

%2) RQNN for nonlinear FAID;

%3) subgraph-expansion-contraction;

%4) FER loss function

%Section IV:

%Case study and numerical results

\vspace{-10pt}
\bibliographystyle{IEEEtran}
\bibliography{./bib/references}
\end{document}

%% file: Introduction.tex
\section{Introduction}
Deep neural networks (DNNs) have gained intensive popularity in communication, signal processing, and data storage communities in the past five years due to their great potential of solving problems relevant to optimization, function approximation, and others.  %\cite{balatsoukas2019deep,Goutay2019,YeTWC2020}. 
One popular idea, known as \emph{model-driven neural networks} (NNs), is to combine the model knowledge (or the prototype algorithms) and the NN in conjunction with optimization techniques of NNs to improve the model. In the context of iterative decoding of error correction codes, the  \emph{deep unfolding} framework \cite{hershey2014deep} is particularly attractive as  it naturally unfolds the decoding iterations over the Tanner graph into a deep neural network.
The activation functions are defined by the prototype decoding rules that mimic the message-passing process \cite{balatsoukas2019deep}. One merit of this framework is that the weight matrices and activation functions over hidden layers are constrained to preserve the message update function symmetry conditions, thus making it possible to train the NN on a single codeword and
its noisy realizations, rather than on the entire code space.

It has been shown that deep unfolding efficiently optimizes various iterative decoding algorithms such as \emph{Belief Propagation} (BP) and improves the decoding convergence  \cite{nachmani2016learning,lugosch2017neural,vasic2018learning,XiaoGC2019,XiaoTCOM2020}.
One common feature of most existing model-driven NNs is that the training sets are constructed by \emph{randomly} sampling channel output sequences at low signal-to-noise-ratios (SNRs), when applied on the additive white Gaussian noise (AWGN) channel 
\cite{nachmani2016learning,lugosch2017neural,vasic2018learning,XiaoGC2019}, or at 
high crossover probabilities in the case of the binary symmetric channel (BSC) \cite{XiaoTCOM2020}. Consequently, when considering the \emph{Low-Density Parity Check} (LDPC) codes, this means that the model-driven NNs are optimized for the waterfall region {{in the curve of decoding performance, where
the probability of error starts to drop drastically}}. %\red{What is a waterfall region? of what plot/curve?} %, which is an abrupt degradation of decoding performance in high SNRs and low crossover probabilities regions \cite{richardson2003error}.  %Furthermore, it is not an easy task to design a single decoder that performs well in both waterfall and error floor regions for a given LDPC code.
However, at low crossover probabilities, low-weight error patterns occur more frequently, and are more likely to cause decoding failure. This requires very long training time to get sufficient statistics on the error floor, making the NN framework impractical for applications that require very small frame error rate (FERs). 

The model-driven NNs are agnostic to these problematic uncorrectable error patterns that dominate the error floor, and to make learning  efficient, it is thereby crucial to  sample from a much smaller set of the most ``harmful'' error patterns. In addition, to optimize the error floor, our NN must use a new loss function based on FER.

It is well known that the error floors of LDPC codes are caused by specific sub-graphs of the Tanner graph, known as \emph{Trapping sets} (TS) \cite{richardson2003error}, which prevent an iterative decoder from converging to a codeword. %away from the maximum likelihood decoding (MLD) for finite-length LDPC codes.  
%In particular, 
When an LDPC code is transmitted over a BSC and is decoded with a specific decoder $\cal D$, the slope of its error floor is characterized by its guaranteed error correction capability $t$ \cite{Ivkovic_TIT_2008}, defined as the largest weight of all error patterns correctable by the decoder $\cal D$, i.e., $\cal D$ can correct all error patterns with weight up to $t$. Generally speaking, the guaranteed error correction capability of a \emph{single} iterative decoder is far lower than that of maximum likelihood decoder (MLD). To increase the guaranteed error correction capability of a given LDPC code, Declercq \emph{et al.} \cite{Declercq_TCOM_2013} proposed an ensemble of finite alphabet iterative decoders (FAIDs), known as \emph{decoder diversity}, where each FAID can correct different error patterns. The FAIDs, if designed properly, are known to be capable of surpassing the floating-point BP algorithms \cite{Planjery2013}. In \cite{Declercq_TCOM_2013}, these decoders are selected  by going over all error patterns in predefined trapping sets, such that their combination can correct all the error patterns associated with the predefined trapping sets. {{Note that this FAID selection is essentially a brute force approach that checks all error patterns for all FAID candidates, and it makes a-priori assumptions on problematic trapping sets, which might not be ``harmful'' trapping sets for specific Tanner graphs}}\cite{Raveendran_TCOM_2020}.

In this paper, we propose a model-driven NN scheme to design the decoder diversity of FAIDs for regular LDPC codes over BSC, with the objective of performing well in both waterfall and error floor regions. {{Unlike the brute force approach in}} \cite{Declercq_TCOM_2013}, {{our framework is dynamically driven by error patterns to design different FAIDs.}} The scheme begins with the design of an initial FAID with a good decoding threshold to guarantee waterfall performance. The rest of the FAIDs are designed via a recurrent quantized neural network (RQNN) in order to reduce the error floor. This RQNN models the universal family of FAIDs, thus it is capable of learning any arbitrary FAID. 
{{To collect sufficient knowledge of the trapping sets, we need to construct a training set consisting of the most problematic error patterns the initial FAID fails on. For this, we rely on the sub-graph expansion-contraction}} \cite{Raveendran_TCOM_2020}.
%To collect sufficient knowledge of the trapping sets, we construct the training set by the sub-graph expansion-contraction \cite{Raveendran_TCOM_2020} to find the most problematic error patterns the initial FAID fails on, rather than taking all error patterns in predefined trapping sets as in \cite{Declercq_TCOM_2013}. 
The advantage is that the sub-graph expansion-contraction method obtains the training set of harmful error patterns for any FAID without making any a-priori assumptions about which graph topologies are harmful. In addition, the method is computationally efficient compared to Monte Carlo simulation and accurate in comparison with other TS enumeration techniques that do not take the decoder into account. 
Instead of selecting the FAIDs by {{checking all error patterns in predefined trapping sets}} as in \cite{Declercq_TCOM_2013}, we train an RQNN on different error patterns to design FAIDs in a sequential fashion. 
%To construct the training set,  rather than taking all error patterns in predefined trapping sets as in \cite{Declercq_TCOM_2013}, we apply the sub-graph expansion-contraction \cite{Raveendran_TCOM_2021} to find the most problematic error patterns the initial FAID fails on.  
Since our goal is to correct specific error patterns rather than reducing the bit error rate (BER), we propose the frame error rate (FER) as the loss function to train the RQNN.  Consequently, the learned FAIDs are optimized in the error floor region and are expected to correct different error patterns. We use the \emph{quasi-cyclic} (QC) Tanner code (155, 64) as an example  \cite{tannercode}, and the numerical results show that the RQNN-aided decoder diversity increases the guaranteed error correction capability {and has a lower error floor bound}. 

The rest of the paper is organized as follows. Section II presents the the necessary preliminaries and the concepts of FAID and decoder diversity. Section III introduces the RQNN framework for designing the decoder diversity. Section IV provides a case study with the Tanner code (155, 64) and demonstrates numerical results. Section V concludes the paper.

%% file: Preliminaries.tex
\section{Preliminaries}
\subsection{Notation}
We consider a binary LDPC code $\cal C$, with a parity-check matrix $\bf H$ of size $M \times N$. The associated Tanner graph is denoted by ${\cal G} = (V,C,E)$, with $V$ (respectively, $C$) the set of $N$ variable (respectively, $M$ check) nodes corresponding to the $N$ columns (respectively, $M$ rows) in $\bf H$, and $E$ the set of edges. Let the $i$-th variable node (VN) be $v_i$, $j$-th check node (CN) be $c_j$, and the edge connecting $v_i$ and $c_j$ be $(v_i, c_j)$ indexed by some integer $(e)$, with $1\le i\le N$, $1\le j\le M$, $1\le e \le |E|$. The set of check (respectively, variable) nodes adjacent to $v_i$ (respectively, $c_j$) is denoted as ${\cal N}(v_i)$ (respectively, ${\cal N}(c_j)$). The degree of a node in $\cal G$ is defined as the number of its neighbors. If all the variable nodes have the same degree $d_v$, $\cal C$ is said to have regular column weight $d_v$, and if all the check nodes have the same degree $d_c$, $\cal C$ is said to have regular row weight $d_c$. In the following, we mainly consider the regular  {$(d_v,d_c)$} LDPC codes. Assume that the crossover probability of BSC is $\alpha$, the codeword transmitted over BSC is ${\bf x} = (x_1,x_2,\cdots, x_N)\in {\cal C}$, and the received channel output vector is ${\bf y}=(y_1,y_2,\cdots,y_N)\in {{\rm GF}(2)}^N$. Let ${\bf e} = (e_1,e_2,\cdots, e_N)$ be the \emph{error pattern} introduced by the BSC, then $\bf y = x \oplus e$, where $\oplus$ is the component-wise XOR operator. The support of an error pattern $\bf e$ is defined as the set of all the locations of nonzero components, namely, ${\rm supp}({\bf e})=\{i:e_i\ne 0\}$, and the weight of an error pattern $\bf e$ denoted by $w({\bf e})$ is defined as the total number of nonzero components, i.e., $w({\bf e})=\left|{\rm supp}({\bf e})\right |$.

A trapping set (TS) $\mathbf{T}$ \cite{richardson2003error, Planjery2013} for an iterative decoder is a non-empty set of variable nodes in $\cal G$ that are not correct at the end of a given number of iterations. A common notation used to denote a TS is $(a,b)$, where $a=|\mathbf{T}|$, and $b$ is the number of odd-degree check nodes in the sub-graph induced by $\mathbf{T}$. Note that $\mathbf{T}$ will depend on the decoder input as well as decoder implementation.

\subsection{FAID}
We follow the definition of FAID introduced in \cite{Planjery2013}. A $b$-bit FAID denoted by ${\cal D}_{FAID}$ is defined by a 4-tuple: ${\cal D}_{FAID} = ({\cal M},{\cal Y},\Phi,\Psi)$, where $\mathcal{M}$ is the domain of the messages passed in FAID defined as $\mathcal{M}=\{0,\pm L_{1},\pm L_{2}...,\pm L_{s}\}$, with $L_{i} \in\mathbb{R}^{+}, 1\le i \le s, s \le 2^{b-1}-1$ and $L_{i}>L_{j}$ if $i>j$. For a message $m \in \mathcal{M}$ associated with $v_i$, its sign represents an estimate of the bit value of $v_i$, namely $v_i=0$ if $m>0$, $v_i=1$ if $m<0$, and $v_i=y_i$ if $m=0$, and its magnitude $|m|$ measures the reliability of this estimate. $\mathcal{Y}$ is the domain of channel outputs. For BSC, $\mathcal{Y}=\{\pm {\rm C}\}$ with some ${\rm C} \in \mathbb{R}^{+}$ as we use the bipolar mapping: $0\to {\rm C}$ and $1\to -{\rm C}$. Let $\mathbf{z}=(z_1,z_2,...z_N)$ be the input vector to a FAID, with $ z_i={(-1)}^{y_i}{\rm C}$, $1\le i \le N$. The functions $\Phi$ and $\Psi$ describe the message update rules of variable nodes and check nodes, respectively. For a check node $c_j$ with degree $d_c$, its updating rule is given by 
\begin{equation}
\Psi ({{\bf{m}}_j}) = \prod\limits_{m \in {{\bf{m}}_j}} {{\mathop{\rm sgn}} (m)}  \cdot \mathop {\min }\limits_{m \in {{\bf{m}}_j}} (|m|),\end{equation}
where ${\rm sgn}(\cdot)$ is the sign function and ${{\bf{m}}_j}$ is the set of extrinsic incoming messages to $c_j$, with $|{{\bf{m}}_j}|=d_c-1$ and ${{\bf{m}}_j} \in \mathcal{M}^{d_c-1}$. For a variable node $v_i$ with degree $d_v$, its updating rule is given by 
\begin{equation}
\Phi ({z_i},{{\bf{n}}_i}) = Q\left( {\sum\limits_{m \in {{\bf{n}}_i}} m  + {\omega _i}{z_i}} \right),
\label{eq:Q}
\end{equation}
where ${{\bf{n}}_i}$ is the set of extrinsic incoming messages to $v_i$,  %and $z_i$ \red{is the bipolar mapping of $v_i$,-- we already defined $z_i$. This statement was a little confusing.}, 
with $|{{\bf{n}}_i}|=d_v-1$ and ${{\bf{n}}_i} \in \mathcal{M}^{d_v-1}$. The function $Q(\cdot)$ is the quantizer defined by $\mathcal{M}$ and a threshold set $\mathcal{T}=\{ T_{1}, ..., T_{s},T_{s+1}=\infty\}$, with $T_{i} \in\mathbb{R}^{+}, 1\le i \le s$ and $T_{i}>T_{j}$ for any $i>j$:
\begin{equation}
Q(x) =
    \begin{cases}
{\mathop{\rm sgn}}(x){L_{i}} & \text{if}\quad T_{i} \le |x| < T_{i + 1}\\
0 & \text{if}\quad |x| < T_{1}
    \end{cases}. %\nonumber
    \label{Q}
    \end{equation}
The coefficient $\omega_i$ is a %non-negative 
real number. If $\omega_i$ is a constant for all possible ${{\bf{n}}_i}$, $\Phi$ is {the quantization of} a linear function, and its associated FAID is called \emph{linear FAID}, otherwise, its associated FAID is called \emph{nonlinear FAID}. 
At the end of each iteration, the estimate of bit associated with each variable node $v_i$ is made by the sign of the sum of all incoming messages and {channel value $z_i$}, i.e., zero if {the} sum is positive, one if {the} sum is negative, and $y_i$ if {the} sum is zero. {This} sum represents the estimate of bit-likelihoods and we denote this operation by $\Upsilon$.

% Since $\Phi$ is a function mapping from ${\cal Y}\times {\cal M}^{d_v-1}$ to ${\cal M}$, for convenience, it can be described as a $d_v-1$-dimensional array or look-up table (LUT). We denote the LUT corresponding to $\Phi$ as $\Phi_v$. Let $\mathcal{M}=\{M_1,M_2,...,M_{2s+1}\}$, with $M_1=-{L_s}, M_2=-{L_{s-1}},...,M_{s}=-L_1, M_{s+1}=0,M_{s+2}=L_1,M_{s+3}=L_2,...,M_{2s+1}=L_{s}$. Then, for the case that the received channel value is $-\rm{C}$, the entry $(i_1,...,i_{d_v-1})$ of $\Phi_v$  
% is computed by 
% \begin{equation}
% {\Phi}_v({i_1},...,{i_{{d_v} - 1}}) = Q\left( { -  \omega_{{i_1},...,{i_{{d_v} - 1}}}\cdot\mathrm{C} + \sum\limits_{j = 1}^{{d_v} - 1} {{M_{{i_j}}}} } \right),
% \label{FAID}
% \end{equation}
% where $1\le i_1,...,i_{d_v-1}\le 2s+1$, and $\omega_{{i_1},...,{i_{{d_v} - 1}}}$ is the $\omega$ corresponding to the entry $(i_1,...,i_{d_v-1})$. Since {$\Phi$} is symmetric, the LUT of the case that the received channel value is $\rm{C}$ can be obtained simply with $-\Phi_v(2s+2-i_1,...,2s+2-{i_{{d_v} - 1}})$.
\subsection{Decoder diversity}
In this work, we follow the definition of decoder diversity in \cite{Declercq_TCOM_2013}. The $b$-bit decoder diversity $\cal D$ is a set consisting of $N_{\cal D}$ $b$-bit FAIDs, which can be defined as 
\begin{equation}
    {\cal D} = \left \{ {\cal D}_i| i = 1,\cdots,N_{\cal D}\right\},
\end{equation}
where each $b$-bit FAID is given by ${\cal D}_i = \left({\cal M},{\cal Y},\Phi_i,\Psi \right)$, with $\Phi_i$ the VN updating rule of ${\cal D}_i$. Given an LDPC code $\cal C$ and a set ${\cal E}_t$ consisting of all error patterns with weight no greater than $t$, the objective is to design $\cal D$ with the smallest cardinality $N_{\cal D}$ such that the $N_{\cal D}$ FAIDs can collectively correct all the error patterns in ${\cal E}_t$. %For a given FAID ${\cal D}_i$, denote the subset of ${\cal E}_t$ that can be corrected by ${\cal D}_i$ as 
The selected FAIDs can be used in either a sequential or a parallel fashion, depending on the memory and throughput constraints. Since each error pattern in ${\cal E}_t$ can be corrected by one or more FAIDs, and each FAID can correct multiple error patterns, this design problem is essentially the Set Covering Problem, which is NP-hard \cite{karp1972reducibility}. In next section, we propose a greedy framework to design $\cal D$ via an RQNN,
which might not have the smallest cardinality $N_{\cal D}$ but still be capable of correcting a great number of error patterns in ${\cal E}_t$.

%% file: Intellgient_Decoder_Diversity.tex
\section{Decoder Diversity by RQNN}
\label{sec:III}
In this section, we formally introduce the design of decoder diversity of FAIDs via recurrent quantized neural networks. For simplicity, we call the decoder diversity of FAIDs as \emph{FAID diversity}.
\subsection{A sequential framework}
\label{sec:III-A}
\begin{figure}
\centering
\includegraphics[width=0.4\textwidth]{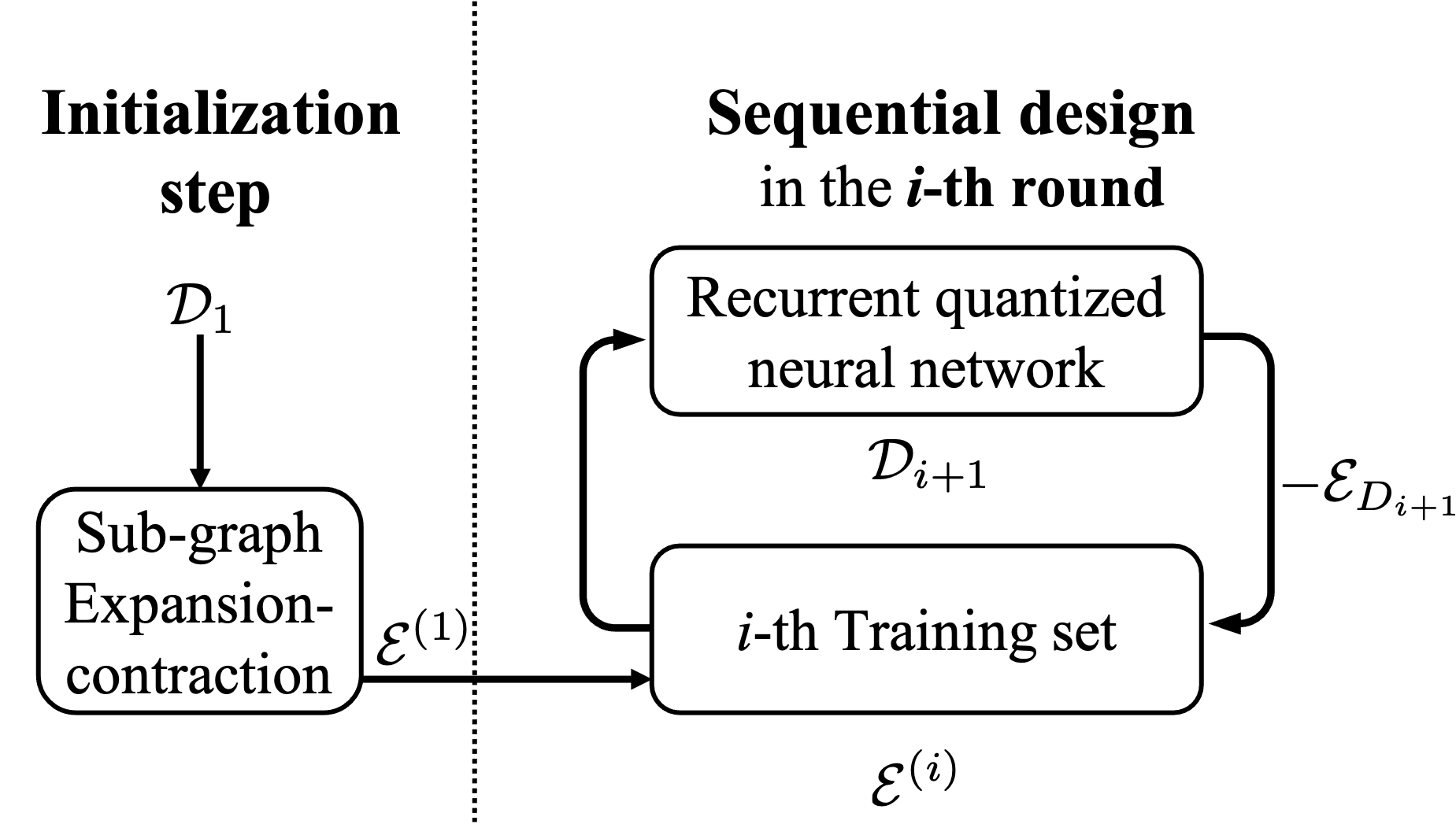}
\caption{Block diagram of FAID diversity via a recurrent quantized neural network. The design process consists of two steps, the Initialization step shown on the left, and the Sequential design step shown on the right. In the Initialization, the sub-graph expansion-contraction is applied for ${\cal D}_1$ to find the set of the most problematic error patterns ${\cal E}^{(1)}$, and ${\cal E}^{(1)}$ is set to be the first training set. In the $i$-th round of Sequential design, the RQNN is trained with the training set ${\cal E}^{(i)}$ to design the FAID  ${\cal D}_{i+1}$. Once the training is completed, ${\cal E}_{{\cal D}_{i+1}}$, the subset of ${\cal E}^{(i)}$ correctable by ${\cal D}_{i+1}$, is excluded from ${\cal E}^{(i)}$.}
\label{fig:block_diagram}
\end{figure}

Basically, the proposed design of $\cal D$ consists of two main steps, namely, the Initialization and Sequential design, as shown in Fig. \ref{fig:block_diagram}. 
In the Initialization step, we begin with the first FAID ${\cal D}_1$ that has a good decoding threshold for the purpose of good performance in waterfall region. ${\cal D}_1$ can be designed by various methods, such as the Density Evolution \cite{nguyen2018analysis} %Information Bottleneck \cite{lewandowsky2018information},
and quantized neural networks \cite{XiaoGC2019,XiaoTCOM2020}. We then use the sub-graph expansion-contraction \cite{Raveendran_TCOM_2020} (as will be introduced in Section \ref{sec:III-B}) to determine the most problematic error patterns that cannot be corrected by ${\cal D}_1$. Denote the set consisting of most problematic error patterns as ${\cal E}^{(1)}$ and the guaranteed correction capability of ${\cal D}_1$ as $t$, {then for any error pattern ${\bf e}\in {\cal E}^{(1)}$, $w({\bf e})=t+1$.} In particular, ${\cal E}^{(1)}$ will be used as the initial training set in the next step.

In the sequential design, we use a recurrent quantized neural network to construct the rest of the FAIDs with the objective of correcting as many error patterns in ${\cal E}^{(1)}$ as possible. We train the RQNN in multiple rounds, with one FAID per round. In each round, the RQNN is trained over a training set consisting of all the error patterns that cannot be corrected by any FAID in the most recently updated set $\cal D$. Once the offline training of the RQNN is completed, the learned FAID corresponding to the current RQNN is added to $\cal D$, and the error patterns that can be corrected by this FAID are excluded from the training set in the current round. 

To be more specific, consider the $i$-th round as shown in Fig. \ref{fig:block_diagram}, where $i\ge 1$. %with ${\cal D}_i$ designed. 
Let the training set used in the $i$-th round be ${\cal E}^{(i)}$, the FAID to be learned in the $i$-th round be ${\cal D}_{i+1}$, and the subset of ${\cal E}^{(i)}$ corrected by the learned FAID be ${\cal E}_{{\cal D}_{i+1}}$. 
Then, for any error pattern ${\bf e}\in {\cal E}^{(i)}$, it cannot be corrected by any of ${\cal D}_1,{\cal D}_2,\cdots,{\cal D}_i$. %The RQNN is trained with ${\cal E}^{(i)}$ and aims to find a ${\cal D}_{i+1}$ such that the cardinality of ${\cal E}_{{\cal D}_{i+1}}$ is maximum. 
The RQNN is trained with ${\cal E}^{(i)}$ {{to minimize the training set  error so that}} the cardinality of ${\cal E}_{{\cal D}_{i+1}}$ is maximum. 
%\red{question can arise how is it ensured? Can we order the $D_i$ in that way ?} 
Subsequently, ${\cal D}_{i+1}$ is added to $\cal D$ and the training set used in the next round ${\cal E}^{(i+1)}$ is derived by 
\begin{equation}
    {\cal E}^{(i+1)} = {\cal E}^{(i)}\setminus {\cal E}_{{\cal D}_{i+1}}.
\end{equation}
This process continues until ${\cal E}^{(i)}$ becomes an empty set or a predefined maximum number of rounds $N_{\cal D}-1$ has been reached. If ${\cal E}^{(i)}$ eventually becomes empty, the designed FAID diversity can correct all error patterns with weight up to $t+1$ in ${\cal E}^{(1)}$. Otherwise, the process have completed $N_{\cal D}-1$ rounds and the last training set ${\cal E}^{(N_{\cal D})}$ is a nonempty set. As a result, the designed FAID diversity can correct most of the error patterns with weight up to $t+1$, and only $\left |{\cal E}^{(N_{\cal D})}\right |$ uncorrectable error patterns with weight $t+1$. 
%The number of error patterns that are not corrected by the FAID diversity is $\left |{\cal E}^{(I_{\max})}\right |$, 

Because of the sequential training strategy, the FAID diversity $\cal D$ uses its decoders in the same order  as how its decoders are determined. Specifically, assume that the predefined maximum number of iterations of ${\cal D}_i$ is $I_{i}$. To decode an input vector $\bf z$, ${\cal D}_1$ with $I_{1}$ iterations is first applied. If ${\cal D}_1$ decodes $\bf z$ successfully, the decoding process terminates, otherwise it is re-initialized with $\bf z$ and switches to ${\cal D}_2$ with $I_{2}$ iterations. 
%If ${\cal D}_2$ decodes $\bf z$ successfully, the decoding process terminates, otherwise the we switch to ${\cal D}_3$ with $I_{3}$ iterations. 
The decoding process continues decoding $\bf z$ with FAIDs in order until a FAID ${\cal D}_j$ corrects $\bf z$. Otherwise, all FAIDs in $\cal D$ fail on $\bf z$, in this case, the decoding process claims a decoding failure. 

%how to use $\cal D$ in actual decoding?
%\red{How to estimate the FER?}
\subsection{Subgraph expansion-contraction}
\label{sec:III-B}
The expansion-contraction method introduced in \cite{Raveendran_TCOM_2020} estimates the error floor of an arbitrary iterative decoder operating on a given Tanner graph of an LDPC code in a computationally efficient way by identifying the minimal-weight uncorrectable error patterns and harmful TSs for the code and decoder. In the expansion step of the method, a list of short cycles (of length $g$ and $g+2$, where $g$ is the girth of the Tanner graph) present in the Tanner graph $\cal G$ of the code is expanded to each of their sufficiently large neighborhood in $\cal G$. The expansion step for a given LDPC code only needs to be executed once and its output: $\mathcal{L}_{\text{EXP}}$, the list of expanded sub-graphs can be used for different decoders for the next step of contraction. 

Each of the expanded sub-graphs in $\mathcal{L}_{\text{EXP}}$ are now \textit{contracted} in order to identify failure inducing sets for a given decoder $\mathcal{D}$. This is achieved by exhaustively decoding error patterns running $\mathcal{D}$ on these sub-graphs (not on the entire Tanner graph $\cal G$) ensuring that the messages accurately represent the actual messages operating on $G$. This process lists all minimum-weight failure inducing error patterns which will then be used as the training set for the RQNN.
%\red{A question from Professor Ravi: can we mention the complexity of this expansion-contraction method?}

The creation of the initial list followed by the expansion procedure represents a negligible portion of the overall computational complexity. The computationally intensive step is the exhaustive contraction of expanded sub-graphs whose complexity can be adjusted based on the number of VNs in the expanded sub-graphs. %Also, for our task of obtaining uncorrectable error patterns for several decoders for the same LDPC code, the cycle counting and the expansion steps need to be performed only once, while the contraction step needs to be reproduced for every decoder. 

\subsection{A model-driven structure for general FAID}
\label{sec:III-C}
\begin{figure}[t]
\centering
\includegraphics[width=3.4in]{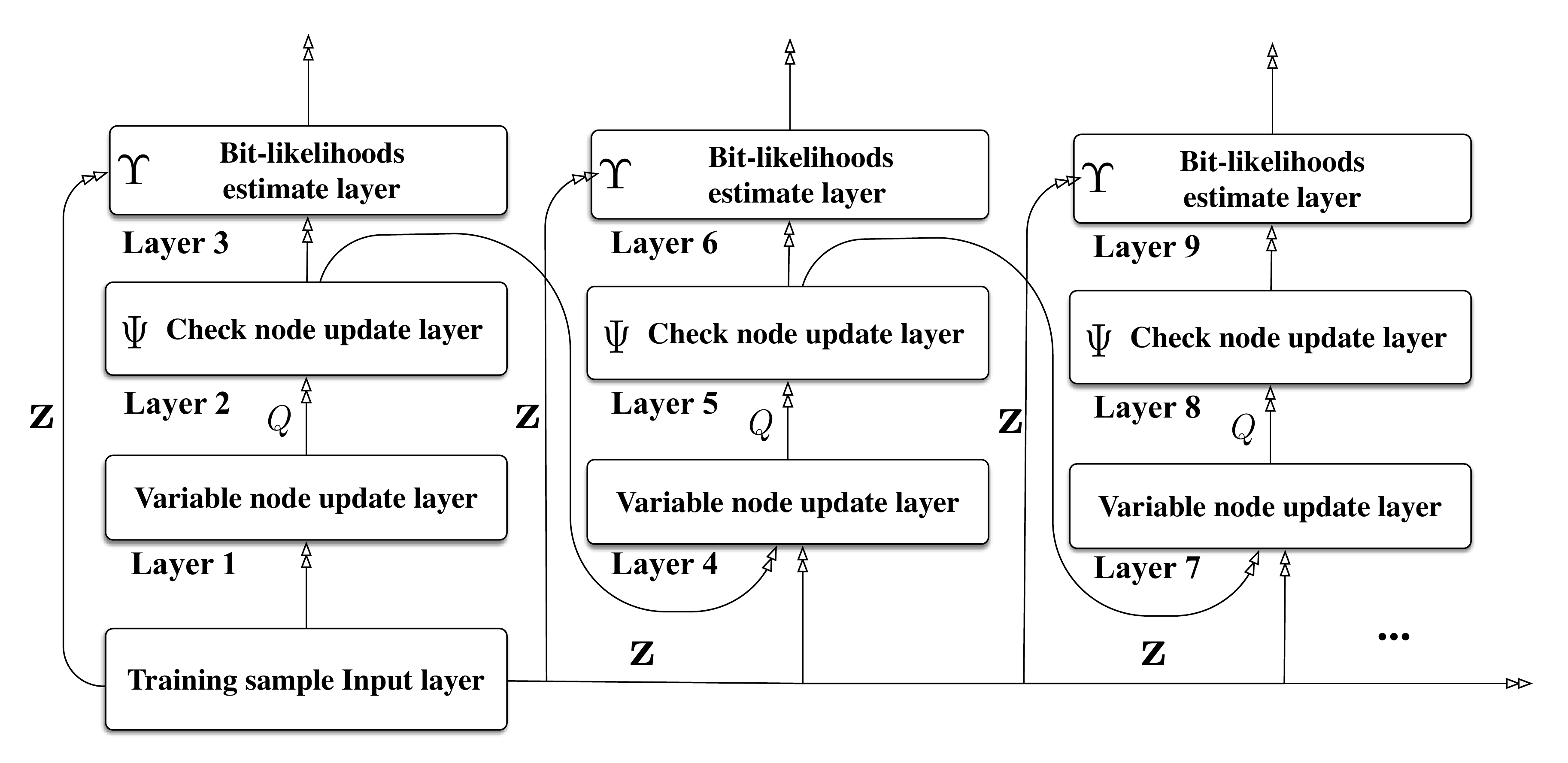}
\caption{Block diagram of a RQNN. Each column corresponds to one iteration, where variable nodes are first updated, followed by the quantization function $Q$. The quantized messages are then used to update check nodes. The output of $\Psi$ and Input layer are fed into both $\Upsilon$ {in current iteration} and the variable node update layer in next iteration.}
\label{fig:RQNNfig}
\end{figure}

As we mentioned in section \ref{sec:III-A}, in the Sequential design, we rely on a recurrent quantized neural network to design ${\cal D}_i$, $i\ge 2$. The proposed RQNN is a model-driven recurrent deep neural network, which is constructed by {unfolding} the general FAID with a given number of iterations, as shown in \cite{nachmani2016learning}. The connection between consecutive layers is determined by the Tanner graph $\mathcal{G}$, and activation functions over hidden layers are defined based on $\Phi$, $\Psi$, and $\Upsilon$. As it has recurrent structure, the trainable parameters are shared among all the iterations. %In particular, rather than introducing weights matrices between layers as in most existing model-driven NN frameworks, the RQNN has only the coefficients $\omega$ in \eqref{eq:Q} as trainable parameters. 
More specifically, suppose that the RQNN consists of $K$ hidden layers, with output values denoted by $\mathbf{r}^{(k)}, 1 \le k \le K$. $\mathbf{r}^{(0)}$ ({respectively}, $\mathbf{r}^{(K+1)}$) represents the values of input ({respectively}, output) layer. $\mathbf{r}^{(k)}=(r_{1}^{(k)},r_{2}^{(k)},...r_{J_k}^{(k)})^T$, where $J_k$ is the number of neurons in $k$-th layer, and $r_{j}^{(k)}$ is the the output value of the $j$-th neuron in $k$-th layer, $0 \le k \le K+1, 1 \le j \le J_k$. As shown in Fig. \ref{fig:RQNNfig}, {three consecutive hidden layers in one column} correspond to one iteration in FAID. For $l \ge 0$, $\mathbf{r}^{(3\ell+1)}$ ({respectively}, $\mathbf{r}^{(3\ell+2)}$) represents the variable (respectively, check) nodes message update, and $\mathbf{r}^{(3\ell)}$ represents the estimate of bit-likelihoods except for $\mathbf{r}^{(0)}$. Each neuron stands for an edge in variable and check nodes message update layers. Each neuron in bit-likelihood approximation layers stands for a variable node. Therefore, we have {$J_k=N$} if $3|k$, otherwise $J_k=|E|$. %Because of the recurrent structure, there are three weight matrices $\mathbf{W}_0$ {(of size $I\times N$)}, $\mathbf{W}_1$ {(of size $I\times I$)} and $\mathbf{W}_2$ {(of size $N\times I$)} used in initialization, variable node update, and bit-likelihood approximation, respectively. Similarly, there are two biases vectors $\mathbf{b}_1$ and $\mathbf{b}_2$ used in variable node update and bit-likelihood approximation, respectively. The $(i,j)$-th entry of $\mathbf{W}_k$ is denoted as $\mathbf{W}_k(i,j)$, and the $(i)$-th entry of $\mathbf{b}_k$ is denoted as $\mathbf{b}_k(i)$. 

In particular, rather than introducing weights matrices between layers as in most existing model-driven NN frameworks, the RQNN has only the coefficients $\omega$ in \eqref{eq:Q} as trainable parameters. Since $\Phi$ is a symmetric function mapping from ${\cal Y}\times {\cal M}^{d_v-1}$ to ${\cal M}$, to design a FAID for a given quantizer $Q(\cdot)$, we only need to determine $\left|{\cal M}\right|^{dv-1}$  values of $\omega$ for the case where the input value is $-\rm C$. 
We use $\omega[i_1,i_2,\cdots,i_{d_v-1}]$ to indicate the coefficient $\omega$ to be learned for the case that the extrinsic incoming message is $(i_1,i_2,\cdots,i_{d_v-1})$, where $i_k \in {\cal M}, k = 1,2,\cdots,d_v-1$.
%We denote the set of these trainable coefficients by $\Omega$, i.e., $\mathrm{\Omega}\buildrel \Delta \over = \left\{ \omega[i_1,i_2,\cdots,i_{d_v-1}]|i_k\in {\cal M}, k = 1,2,\cdots,d_v-1\right\}$, where $\omega[i_1,i_2,\cdots,i_{d_v-1}]$ indicates the coefficient $\omega$ to be learned for the case that the extrinsic incoming message is $(i_1,i_2,\cdots,i_{d_v-1})$.
Moreover, we are interested in the $\Phi$ that is invariant to the ordering of the extrinsic incoming messages. This means that $\Phi ({z_i},{{\bf{n}}_i}') = \Phi ({z_i},{{\bf{n}}_i})$, where ${{\bf{n}}_i}'$ is an arbitrary permutation of ${{\bf{n}}_i}$. This rotation symmetry of $\Phi$ limits the RQNN to learn the coefficients with non-decreasing arguments, namely,  $\omega[i_1,i_2,\cdots,i_{d_v-1}]$ where $i_j\le i_k, \forall j<k$, 
and reduces the number of trainable coefficients from $\left|{\cal M}\right|^{dv-1}$ to $\binom{|{\cal M}|+d_v-2}{d_v-1}$. We call the FAID whose $\Phi$ has rotation symmetry property as \emph{symmetric} FAID. 
We denote the set of these trainable coefficients by $\Omega$, i.e., $\mathrm{\Omega}\buildrel \Delta \over = \left\{ \omega[i_1,i_2,\cdots,i_{d_v-1}]|i_k\in {\cal M}, i_1\le i_2\le \cdots \le i_{d_v-1}\right\}$, and $\mathop{\cal P}({\bf n})$ be the sorted vector of the set ${\bf n}$ in non-decreasing order. 

Then, the output of the $k$-th hidden layer ($k>1$) is computed as follows:
\begin{eqnarray}
\mathbf{r}^{(k)}=
    \begin{cases}
      \Phi\left(\Omega,\mathbf{z}^{\rm T},\mathbf{r}^{(k-2)}\right), &3|(k -1) \\
			\Psi\left(\mathbf{r}^{(k-1)}\right), &3|(k-2)\\
			\Upsilon\left(\mathbf{z}^{\rm T}, \mathbf{r}^{(k-1)}\right), &3|k
    \end{cases}, %\nonumber
    \label{eq:acf}
\end{eqnarray}
where for any $(e) = ({v_i},{c_j})$, if $3|(k-1)$,
\begin{eqnarray*}
{r_{e}^{(k)}}=
    \begin{cases}
      Q\left( {\sum\limits_{m \in {{\bf{n}}}} m  + \omega\left[\mathop{\cal P}({\bf n})\right]{z_i}} \right), &z_i = -{\rm C} \\
			-Q\left( {\sum\limits_{m \in {{\bf{n}'}}} m  + \omega\left[\mathop{\cal P}({\bf n}')\right]{z_i}} \right), &z_i = +{\rm C}
    \end{cases}, %\nonumber
    %\label{eq:acf}
\end{eqnarray*}
with ${\bf n} = \left \{  r_{e'}^{(k-2)}:(e') = ({v_i},{c_{j'}}),j' \ne j\right \}$, and ${\bf n}'=\{-m:m\in {\bf n}\}$.

% \begin{equation*}
%     %\begin{array}{*{20}{l}}
% {r_{e}^{(k)}} =
% Q\left( \sum\limits_{\substack{(e') = ({v_i},{c_{j'}})\\j' \ne j}} {{r_{e'}^{(k-2)}}} +\omega\left[\mathop{\cal P}\limits_{\substack{(e') = ({v_i},{c_{j'}})\\j' \ne j}}{{r_{e'}^{(k-2)}}}\right]z_i \right),
% %\end{array},
% \end{equation*}

% \begin{eqnarray*}
% {r_{e}^{(k)}} = &\\
% Q\left( \sum\limits_{(e') = ({v_i},{c_j'}),j' \ne j} {{r_{e'}^{(k-2)}}} +\omega\left[\mathop{\rm Perm}\limits_{(e') = ({v_i},{c_j'}),j' \ne j}{{r_{e'}^{(k-2)}}}\right]z_i \right),
% \end{eqnarray*}
if $3|(k-2)$, 
\[{r_{e}^{(k)}} = \prod\limits_{\substack{(e') = ({v_{i'}},{c_j})\\i' \ne i}} {{\rm{sgn}}\left({r_{e'}^{(k-1)}}\right)}  \cdot \mathop {\min }\limits_{\substack{(e') = ({v_{i'}},{c_j})\\i' \ne i}} \left| {{r_{e'}^{(k-1)}}} \right|,\]
and if $3|k$, 
\[{r_{i}^{(k)}} =  \sum\limits_{(e') = ({v_i},{c_t})} {{r_{e'}^{(k-1)}}} + z_i.\]
The first hidden layer is the initialization of the RQNN, and its output is calculated by ${r_{e}^{(1)}} = {r_{i}^{(0)}},\forall (e) = ({v_i},{c_j})$. At each $k$-th layer where $3|k$ (i.e., the bit-likelihood estimate layer), for each training sample, we check whether its estimated codeword satisfies all parity equations or not. If so, this training sample will skip the remaining subsequent layers and its output at the current $k$-th layer will be directly used to calculate the {loss} function, namely, ${{\bf{r}}^{(K + 1)}} = {{\bf{r}}^{(k)}},{\text{    if    }}\frac{1}{2}\left( {{\bf{1}} - {\rm{sgn}}({{({{\bf{r}}^{(k)}})}^{\rm T}})} \right) \cdot {{\bf{H}}^{\rm T}} = {\bf{0}}$. If there is no intermediate bit-likelihood estimate layer satisfying all parity equations, the last bit-likelihood estimate layer will be used to compute the {loss} function. Note that the $Q(\cdot)$ function in %\eqref{eq:Q} and 
\eqref{eq:acf} is %are 
predefined, and the CN update layer and bit-likelihood layer are the same as the min-sum (MS) decoder. Therefore, the design of a FAID directly comes from learning the VN update layer.

In \cite{XiaoTCOM2020}, we proposed using RQNNs to design linear FAIDs. This approach, however, has limitations when an optimal FAID is nonlinear whose VN updating rule satisfies specific constraints. In \cite{Planjery2013}, Lemma 1 provided an example of possible constraints and showed that any FAID whose $\Phi$ satisfying these constraints cannot be expressed as a quantization of a linear function. The main reason of these limitations lies in its prototype VN updating rule, which assumes that the coefficient $\omega$ is a constant. On the contrary, the prototype VN updating rule in our proposed RQNN can express any arbitrary symmetric FAID, as shown in the following proposition.
\begin{proposition}
Given an arbitrary symmetric FAID, whose VN updating mapping is determined by $\binom{|{\cal M}|+d_v-2}{d_v-1}$ values $\left\{ \mu[i_1,i_2,\cdots,i_{d_v-1}]|i_k\in {\cal M}, i_1\le i_2\le \cdots \le i_{d_v-1} \right\}$, where $\mu[i_1,i_2,\cdots,i_{d_v-1}] \in {\cal M}$ is the value for the case that the received input is $-\rm C$ and the extrinsic incoming message is $(i_1,i_2,\cdots,i_{d_v-1})$. Then, there exist $\binom{|{\cal M}|+d_v-2}{d_v-1}$ coefficients %real numbers $\left\{ \omega[i_1,i_2,\cdots,i_{d_v-1}]|i_k\in {\cal M}, i_1\le i_2\le \cdots \le i_{d_v-1} \right\}$ 
such that for any $i_k\in {\cal M}$, $i_1\le i_2\le \cdots \le i_{d_v-1}$, $\Phi(-{\rm C},i_1, i_2, \cdots, i_{d_v-1})=\mu[i_1,i_2,\cdots,i_{d_v-1}]$.
\end{proposition}
\beginofproof
First we notice that these $\binom{|{\cal M}|+d_v-2}{d_v-1}$ coefficients are defined independently, thus we can determine each coefficient individually. Consider the extrinsic incoming message $(i_1,i_2,\cdots,i_{d_v-1})$. For simplicity, we use $\mu$ (respectively, $\omega$) to indicate $\mu[i_1,i_2,\cdots,i_{d_v-1}]$ (respectively, $\omega[i_1,i_2,\cdots,i_{d_v-1}]$), and let  $s={\sum\limits_{k=1}^{d_v-1} i_k}$. 
If $\mu = 0$, then
\begin{equation}
\left|s-{\omega}{\rm C}\right|<T_1, \Rightarrow  \frac{\left(s-T_1\right)}{\rm C}<\omega <\frac{\left(s+T_1\right)}{\rm C}.
\label{eq:prop1}
\end{equation}
For $\mu \ne 0$, we have ${\rm sgn\left(s-{\omega}{\rm C}\right)}={\rm sgn}(\mu)$. Let $|\mu| = L_j$ for some $j$. If $\mu>0$, then
\begin{equation}
T_j\le s-\omega{\rm C}<T_{j+1}, \Rightarrow   \frac{\left(s-T_{j+1}\right)}{\rm C}<\omega\le \frac{\left(s-T_j\right)}{\rm C}.
\label{eq:prop2}
\end{equation}
If $\mu<0$, then
\begin{equation}
T_j\le \omega{\rm C}-s<T_{j+1}, \Rightarrow   \frac{\left(s+T_{j}\right)}{\rm C}\le\omega< \frac{\left(s+T_{j+1}\right)}{\rm C}.
\label{eq:prop3}
\end{equation}
Consequently, the necessary and sufficient condition to express the given FAID by the prototype $\Phi$ is to take each coefficient $\omega$ in the ranges specified above. 
%In particular, we can derive a range of each coefficient such that the prototype $\Phi$ can express this given FAID. Consider the extrinsic incoming message  $(i_1,i_2,\cdots,i_{d_v-1})$. For simplicity, we use $\mu$ (respectively, $\omega$) to indicate $\mu[i_1,i_2,\cdots,i_{d_v-1}]$ (respectively, $\omega[i_1,i_2,\cdots,i_{d_v-1}]$). The necessary and sufficient condition for $\omega$ to have $\Phi(-{\rm C},i_1, i_2, \cdots, i_{d_v-1})=\mu$ is
\endofproof

Proposition 1 indicates that our RQNN framework, if properly initialized, can learn any arbitrary FAID.

When the offline training is completed, we obtain a trained $\Omega$, based on which we derive a look-up table (LUT) which is used to describe $\Phi$ and can be efficiently implemented in hardware. In the initialization, the channel input $+\rm{C}$ (respectively, $-\rm{C}$) is first mapped to $+L_1$ (respectively, $-L_1$). For the case that the received channel input is $-\rm{C}$, the LUT used to describe $\Phi$ for all variable nodes  
is a $(d_{v}-1)$-{dimensional} array, denoted as $\Phi_{RQNN}$. Let $\mathcal{M}=\{M_1,M_2,...,M_{2s+1}\}$, with $M_1=-{L_s}, M_2=-{L_{s-1}},...,M_{s}=-L_1, M_{s+1}=0,M_{s+2}=L_1,M_{s+3}=L_2,...,M_{2s+1}=L_{s}$. Then, for $1\le i_1,i_2,...,i_{d_v-1}\le 2s+1$, the entry $(i_1,i_2,...,i_{d_v-1})$ of $\Phi_{RQNN}$
is computed by 
\begin{equation}
\begin{array}{*{20}{l}}
{\Phi}_{RQNN}({i_1},...,{i_{{d_v} - 1}}) =\\
Q\left( { -  \omega[{i_1},...,{i_{{d_v} - 1}}]\cdot\mathrm{C} + \sum\limits_{j = 1}^{{d_v} - 1} {{M_{{i_j}}}} } \right).
\end{array}
\label{eq:FAID}
\end{equation}
Since {$\Phi_{RQNN}$} is symmetric, the LUT of the case that the received channel value is $\rm{C}$ can be obtained simply with $-\Phi_{RQNN}(2s+2-i_1,2s+2-i_2,...,2s+2-{i_{{d_v} - 1}})$.
 %, and its output is calculated by 
% \begin{equation}
% {{\bf{r}}^{(1)}} = Q({{\bf{r}}^{(0)}}),
% \label{ini}
% \end{equation} 
%where ${\bf{r}}^{(0)}=\mathbf{z}^T={(-1)}^{\mathbf{y}^T}\cdot \rm{C}$, ${r_{e}^{(1)}} = Q(\mathbf{W}_0(e,i){r_{i}^{(0)}}),\forall (e) = ({v_i},{c_j})$.

% \subsection{From RQNN to a FAID}
% Since $\Phi$ is a function mapping from ${\cal Y}\times {\cal M}^{d_v-1}$ to ${\cal M}$, for convenience, it can be described as a $d_v-1$-dimensional array or look-up table (LUT). We denote the LUT corresponding to $\Phi$ as $\Phi_v$. Let $\mathcal{M}=\{M_1,M_2,...,M_{2s+1}\}$, with $M_1=-{L_s}, M_2=-{L_{s-1}},...,M_{s}=-L_1, M_{s+1}=0,M_{s+2}=L_1,M_{s+3}=L_2,...,M_{2s+1}=L_{s}$. Then, for the case that the received channel value is $-\rm{C}$, the entry $(i_1,...,i_{d_v-1})$ of $\Phi_v$  
% is computed by 
% \begin{equation}
% {\Phi}_v({i_1},...,{i_{{d_v} - 1}}) = Q\left( { -  \omega_{{i_1},...,{i_{{d_v} - 1}}}\cdot\mathrm{C} + \sum\limits_{j = 1}^{{d_v} - 1} {{M_{{i_j}}}} } \right),
% \label{FAID}
% \end{equation}
% where $1\le i_1,...,i_{d_v-1}\le 2s+1$, and $\omega_{{i_1},...,{i_{{d_v} - 1}}}$ is the $\omega$ corresponding to the entry $(i_1,...,i_{d_v-1})$. Since {$\Phi$} is symmetric, the LUT of the case that the received channel value is $\rm{C}$ can be obtained simply with $-\Phi_v(2s+2-i_1,...,2s+2-{i_{{d_v} - 1}})$.
\subsection{Training RQNN}
Denote the RQNN by $\cal H$. The coefficients $\omega$ in $\cal H$ can be initialized by conventional iterative decoders or the decoders constructed by specific techniques such as Density Evolution and the selection approach in \cite{Planjery2013}. Specifically, we first derive the LUT of some well-known decoder like quantized Offset MS decoder, and take its coefficients as the initialization of $\cal H$. As shown in Proposition 1, there are plenty of different coefficients corresponding to a given LUT. 
Since the RQNN preserves the symmetry conditions,  we can simply use the all-zero codeword and its noisy realizations to construct a training set. In particular, to design ${\cal D}_{i+1}$ in the $i$-th round, the RQNN is trained with ${\cal E}^{(i)}$, namely, ${\bf{r}}^{(0)}=\mathbf{e}^{\rm T}$ where ${\bf e}\in {\cal E}^{(i)}$. Let  $\mathbf{u}=\mathbf{r}^{(K+1)}$ be the values in the output layer, then, $\mathbf{u}=\mathcal{H}\left(\mathbf{r}^{(0)}\right)$, with ${\bf{r}}^{(0)}$ receiving the error patterns in ${\cal E}^{(i)}$. 

Since the objective of $\cal H$ is to correct the error patterns in ${\cal E}^{(i)}$ as many as possible, and we assume $\bf x = 0$, we propose the following frame error rate as the loss function
\begin{equation}
\Gamma ({{\bf{u}}}) = \frac{1}{2}\left[1-{\rm sgn}\left(\mathop {\min }\limits_{1\le i \le N} { u_i}\right)\right].
%\frac{1}{N}\sum\limits_{i = 1}^N {{{\left( {{x_i} -  {{\hat x_i}}} \right)}^2}} ,
\label{eq:loss}
\end{equation}
Note that $u_i$ represents the estimate likelihood of the $i$-th bit. Since $\bf x = 0$, each component in $\bf u$ is expected to be positive. Therefore, a frame is decoded in error if and only if the minimum component is negative. The quantizer $Q(\cdot)$ in \eqref{eq:acf} and the sign function in \eqref{eq:loss} cause a critical issue that their gradients vanish almost everywhere, making it difficult to use classical backward propagation. Similar to \cite{XiaoTCOM2020}, we leverage straight-through estimators (STEs) as surrogate gradients to tackle this issue. In particular, we use the following two STEs for $Q$ and the sign function
\begin{equation}
f_{\cal M}(x) = \left\{ {\begin{array}{*{20}{l}}
1&{{\rm{if}}\quad |x| < {T_{s}}}\\
0&{{\rm{otherwise}}}
\end{array},} \right.
\label{eq:ste1}
\end{equation}
\begin{equation}
f_{\rm sgn}(x) = \frac{{2{e^{ - x}}}}{{{{(1 + {e^{ - x}})}^2}}}.
\label{eq:ste2}
\end{equation}
In the backward propagation, $f_{\cal M}(x)$ and $f_{\rm sgn}(x)$ are used to replace the zero gradients of $Q$ and the sign function, respectively. Motivated by the fact that applying different learning rates to $\Omega$ can help training convergence, we employ ADAM\cite{kingma2014adam} with mini-batches as optimizer, with
gradients accumulated in full precision. 
%Initialization of coefficients
\label{sec:III-D}

%% file: Results.tex
\section{A Case study and numerical results}
\begin{figure}[t]
\centering
\includegraphics[width=3.in]{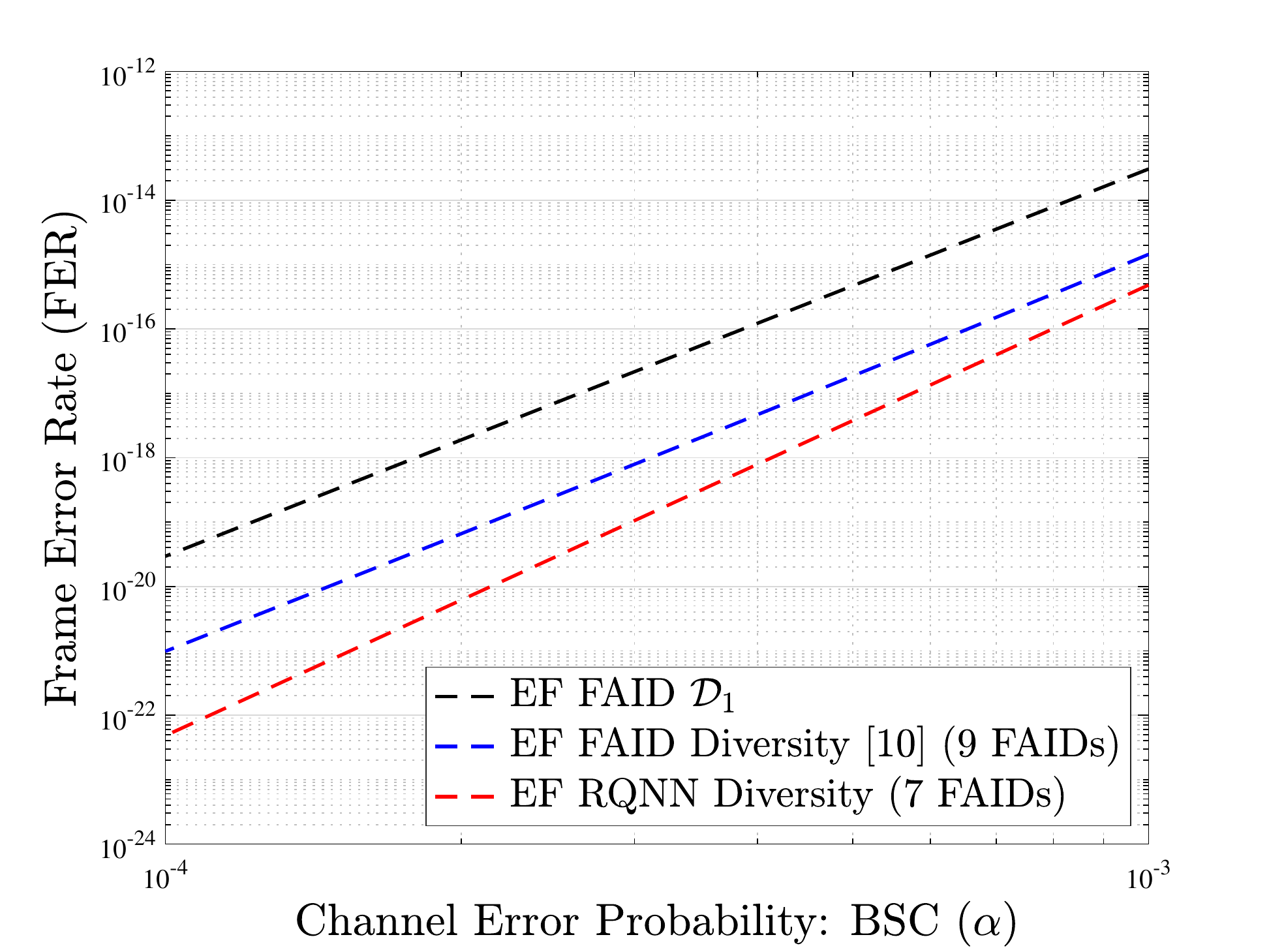}
\caption{Error floor estimation of ${\cal D}_1$, FAID Diversity \cite{Declercq_TCOM_2013}, and RQNN-aided FAID Diversity. FAID Diversity \cite{Declercq_TCOM_2013} has 9 FAIDs with 450 iterations, while the RQNN-aided FAID Diversity has only 7 FAIDs with 390 iterations.}
\label{fig:FER}
\end{figure}

In this section, we consider the Tanner code (155, 64) as an example to demonstrate how to construct an RQNN to design a 3-bit FAID diversity. The Tanner code has column weight of 3 and row weight of 5, with a minimum distance of 20. The $\cal M$ and $\cal T$ are predefined to be ${\cal M}=\{0,\pm 1, \pm 2, \pm 3\}$ and ${\cal T}=\{\pm 0.5, \pm 1.5, \pm 2.5\}$, respectively. We first select the 3-bit nonlinear FAID ${\rm D}_0$ in \cite{Planjery2013}[TABLE II] as our ${\cal D}_1$. As shown in \cite{Planjery2013}, ${\cal D}_1$ is guaranteed to correct all error patterns with weight up to 5. We apply the sub-graph expansion-contraction method for ${\cal D}_1$ with 100 iterations, and obtain ${\cal E}^{(1)}$ consisting of 29294 error patterns with weight of 6 uncorrectable by ${\cal D}_1$. We construct an RQNN with 50 iterations. The size of each mini-batch is set to 20, and the learning rate is set to 0.001. We sequentially train the RQNN in 6 rounds, with the initialization shown in Table \ref{tbl:Initial}. In particular, for each initialization, we start from some FAID and randomly sample its coefficients in the ranges specified in \eqref{eq:prop1}-\eqref{eq:prop3}. For example, we take ${\cal G}_0$ and ${\cal G}_3$ in \cite{Planjery2013} as the initial points for ${\cal D}_2$ and ${\cal D}_3$, respectively. The training of RQNN in 6 rounds converged within 10, 10, 20, 60, 60, 30 epochs, respectively. 
 \begin{table}[t]%[H]
\caption{Initialization of RQNN in 6 rounds for Tanner code (155, 64).}
%\centering
\label{tbl:Initial}
%\begin{adjustbox}{angle=90}
\resizebox{0.48\textwidth}{!}{
\begin{tabular}{|c|c|c|c|c|c|c|}         \hline\hline
$\Omega$ & ${\cal D}_2$ & ${\cal D}_3$ & ${\cal D}_4$ & ${\cal D}_5$ & ${\cal D}_6$ & ${\cal D}_7$ \\ \hline 
$\omega[{-3,-3}]$ & -1.4949 & 0.0436 & -0.0116 & 0.0160 & 0.0180 & 0.0081  \\ \hline
$\omega[{-3,-2}]$ & -1.1411 & 0.0165 & 0.0011 & 0.0027 & -0.0017 & 0.0114  \\ \hline
$\omega[{-3,-1}]$ & -0.3906 & 0.1579 & -0.0013 & 0.0259 & 0.0075 & 0.0117  \\ \hline
$\omega[{-3,0}]$ & 0.6000 & 0.0107 & 0.0152 & -0.0007 & 0.0164 & 0.0128  \\ \hline
$\omega[{-3,1}]$ & 1.5927 & 1.0248 & 1.0090 & 0.9971 & 1.0144 & 1.0101  \\ \hline
$\omega[{-3,2}]$ & 1.9867 & 2.5106 & 2.5018 & 2.4959 & 2.5034 & 2.5076  \\ \hline
$\omega[{-3,3}]$ & 0.9138 & 1.0230 & 0.5055 & 0.9879 & 0.0135 & 1.0200   \\ \hline\hline
$\omega[{-2,-2}]$ & -0.6916 & 0.0238 & 0.0118 & 0.0058 & 0.0252 & 0.0005  \\ \hline
$\omega[{-2,-1}]$ & -0.1420 & 0.0314 & 0.0006 & 0.0052 & 0.0172 & 0.0224  \\ \hline
$\omega[{-2,0}]$ & 0.5015 & 1.1020 & 0.9856 & 1.0005 & 1.0116 & 0.9960  \\ \hline
$\omega[{-2,1}]$ & 1.2733 & 0.5002 & 0.5096 & 0.9983 & 2.4933 & 1.4962  \\ \hline
$\omega[{-2,2}]$ & 0.9185 & 1.0360  & 1.0105 & 1.0055 & 0.9946 & 0.9986  \\ \hline
$\omega[{-2,3}]$ & -0.5289  & 0.0100 & 0.0019 & 0.0101 & -0.0024 & 0.0198  \\ \hline\hline
$\omega[{-1,-1}]$ & 0.1838 & 0.0217  & 0.0011 & -0.0053 & -0.0091 & -0.0052 \\ \hline
$\omega[{-1,0}]$ & 0.5011 & 1.0330 & 1.0167 & 0.4787 & 0.5132 & 0.9917  \\ \hline
$\omega[{-1,1}]$ & 0.9608 & 1.0298 & 1.0239 & 1.00097 & 0.9901 & 1.0001  \\ \hline
$\omega[{-1,2}]$ & 0.6858 & 2.0016 & 1.9976 & 0.0544 & 1.0084 & 2.0008  \\ \hline
$\omega[{-1,3}]$ & -0.0430 & 0.5010 & 0.0419 & 0.5073 & 0.5136 & 1.0247  \\ \hline\hline
$\omega[{0,0}]$ & 0.6353  & 1.0109 & 0.9769 & 1.0010 & 0.9783 & 0.9929  \\ \hline
$\omega[{0,1}]$ & 1.3468 & 1.1060 & 0.9840 & 0.9937 & 0.9976 & 2.0139  \\ \hline
$\omega[{0,2}]$ & 1.5708 &  2.0549  & 2.0066 & 1.0020 & 1.9932 & 2.0050  \\ \hline
$\omega[{0,3}]$ & 1.1832  & 1.0263 & 0.9961 & 1.0098 & 1.5074 & 0.5052 \\ \hline\hline
$\omega[{1,1}]$ & 1.6955  & 2.0526 & 2.0093 & 2.0019 & 1.0550 & 2.0039  \\ \hline
$\omega[{1,2}]$ & 2.2605 & 2.0788  & 2.0147 & 2.0095 & 2.0019 & 1.9809 \\ \hline
$\omega[{1,3}]$ & 1.5015 & 2.1045 & 1.9706 & 1.9975 & 1.9900 & 1.5062   \\ \hline\hline
$\omega[{2,2}]$ & 2.6440 & 3.1120 & 3.0082 & 3.0039 & 1.5147 & 2.5085  \\ \hline
$\omega[{2,3}]$ & 2.5011 & 3.0811 & 2.9859 & 2.9898 & 2.5138 & 1.0195  \\ \hline\hline
$\omega[{3,3}]$ & 3.0139 & 2.0509 & 2.0041 & 1.9976 & 1.9922 & 1.9947  \\ \hline
\end{tabular}}
      % \end{adjustbox}
    \end{table}  
The trained coefficients are provided in Table \ref{tbl:Trained}, from which we derive 6 RQNN-aided FAIDs. Consequently, our RQNN Diversity consists of 7 FAIDs. Set the maximum number of iterations of these 7 FAIDs to $100,90,50,40,50,30,30$ %$100,100,50,50,50,50,50$ 
accordingly. For comparison, we consider the FAID Diversity in \cite{Declercq_TCOM_2013} consisting of 9 FAIDs, with each FAID performing 50 iterations. Note that the FAID Diversity in \cite{Declercq_TCOM_2013} starts from ${\cal D}_1$ as well. Table \ref{tbl:error_pttrns} summarizes the statistics on the error correction of  the FAID Diversity derived by RQNN and the FAID Diversity in \cite{Declercq_TCOM_2013}. The ${\cal E}'$ consists of all 1,147,496 error patterns with weight of 7 uncorrectable by ${\cal D}_1$, which is obtained by applying the sub-graph expansion-contraction method for ${\cal D}_1$ with 100 iterations. The numbers in Table \ref{tbl:error_pttrns} indicate how many error patterns uncorrectable by the corresponding FAID diversity. As shown in Table \ref{tbl:error_pttrns}, the FAID Diversity derived via RQNNs has less uncorrectable error patterns than the FAID Diversity in \cite{Declercq_TCOM_2013}. In particular, the error correction capability of the FAID Diversity in \cite{Declercq_TCOM_2013} is 6, while the error correction capability of the FAID Diversity via RQNNs is 7. 
Fig. \ref{fig:FER} shows the error floor estimation of the FER performance of single FAID ${\cal D}_1$, FAID Diversity in \cite{Declercq_TCOM_2013}, and the FAID Diversity via RQNNs. The bounds of error floor is estimated by the statistics in Table .\ref{tbl:error_pttrns}. As shown in Fig. \ref{fig:FER}, the FAID Diversity via RQNNs has the lowest error floor bound compared to the others.

%Fig. \ref{fig:FER} shows the FER decoding performance of single FAID ${\cal D}_1$, FAID Diversity in \cite{Declercq_TCOM_2013}, and the FAID Diversity via RQNNs. As shown in Fig. \ref{fig:FER}, the FAID Diversity via RQNNs has lower error floor compared to ${\cal D}_1$ and performs almost the same as the FAID Diversity \cite{Declercq_TCOM_2013} in the waterfall region with less FAIDs. 
%Note that the FAID Diversity in \cite{Declercq_TCOM_2013} starts from ${\cal D}_1$ as well. %Table \ref{tbl:error_pttrns} summarizes the  statistics on the error correction of  the FAID Diversity derived by RQNN and the FAID Diversity in \cite{Declercq_TCOM_2013}. Note that the FAID Diversity in \cite{Declercq_TCOM_2013} contains 9 FAIDs and starts from ${\cal D}_1$ as well. 

 \begin{table}[t]%[H]
\caption{Trained coefficients in 6 rounds for Tanner code (155, 64).}
%\centering
\label{tbl:Trained}
%\begin{adjustbox}{angle=90}
\resizebox{0.48\textwidth}{!}{
\begin{tabular}{|c|c|c|c|c|c|c|}         \hline\hline
$\Omega$ & ${\cal D}_2$ & ${\cal D}_3$ & ${\cal D}_4$ & ${\cal D}_5$ & ${\cal D}_6$ & ${\cal D}_7$ \\ \hline 
$\omega[{-3,-3}]$ & -1.4994 & 0.0400 & 0.0031 & 0.0076 & 0.0228 & 0.0038  \\ \hline
$\omega[{-3,-2}]$ & -1.1455 & 0.0105 & 0.0156 & -0.0105 & -0.0094 & 0.0078  \\ \hline
$\omega[{-3,-1}]$ & -0.3951 & 0.1521 & 0.0135 & 0.03830 & -0.0134 & 0.0169  \\ \hline
$\omega[{-3,0}]$ & 0.5944 & 0.0060 & 0.0301 & -0.0231 & 0.0001 & 0.0173  \\ \hline
$\omega[{-3,1}]$ & 1.5881 & 1.0217 & 1.0187 & 1.0113 & 0.9995 & 1.0086  \\ \hline
$\omega[{-3,2}]$ & 1.9813 & 2.5173 & 2.5157 & 2.4718 & 2.5267 & 2.5234 \\ \hline
$\omega[{-3,3}]$ & 0.9079 & 1.0180 & 0.4953 & 0.9479 & 0.0135 & 1.0070   \\ \hline\hline
$\omega[{-2,-2}]$ & -0.6959 & 0.0172 & 0.0263 & 0.0241 & 0.0421 & -0.0006  \\ \hline
$\omega[{-2,-1}]$ & -0.1462 & 0.0270 & 0.0150 & -0.0300 & 0.0226 & 0.0282  \\ \hline
$\omega[{-2,0}]$ & 0.4955 & 1.1009 & 0.9977 & 0.9629 & 0.9939 & 1.0000  \\ \hline
$\omega[{-2,1}]$ & 1.2686 & 0.4989 & 0.4972 & 1.0039 & 2.5120 & 1.5033  \\ \hline
$\omega[{-2,2}]$ & 0.9133 & 1.0319 & 1.0113 & 0.9945 & 0.9760 & 1.0119  \\ \hline
$\omega[{-2,3}]$ & -0.5346 & 0.0169 & -0.0085 & 0.0157 & -0.0241 & 0.0360 \\ \hline\hline
$\omega[{-1,-1}]$ & 0.1794 & 0.0162 & 0.0158 & 0.0066 & 0.0100 & -0.0010 \\ \hline
$\omega[{-1,0}]$ & 0.4978 & 1.0386 & 1.0296 & 0.5281 & 0.4900 & 1.0093  \\ \hline
$\omega[{-1,1}]$ & 0.9568 & 1.0281 & 1.0381 & 0.9634 & 0.9766 & 0.9949  \\ \hline
$\omega[{-1,2}]$ & 0.6858 & 2.0043 & 1.9861 & 0.0125 & 1.0084 & 2.0126  \\ \hline
$\omega[{-1,3}]$ & -0.0395 & 0.4996 & 0.0479 & 0.4955 & 0.4912 & 1.0067  \\ \hline\hline
$\omega[{0,0}]$ & 0.6304  & 1.0147 & 0.9931 & 1.0352 & 0.9542 & 0.9907  \\ \hline
$\omega[{0,1}]$ & 1.3468 & 1.1060 & 0.9840 & 0.9937 & 0.9976 & 1.9892  \\ \hline
$\omega[{0,2}]$ & 1.5708 &  2.0549  & 2.0066 & 1.0213 & 1.9932 & 2.0050  \\ \hline
$\omega[{0,3}]$ & 1.1879  & 1.0304 & 1.0071 & 0.9715 & 1.4905 & 0.4925 \\ \hline\hline
$\omega[{1,1}]$ & 1.6955  & 2.0526 & 2.0093 & 2.0019 & 1.0328 & 2.0039  \\ \hline
$\omega[{1,2}]$ & 2.2563 & 2.0854  & 2.0132 & 1.9881 & 1.9856 & 1.9624 \\ \hline
$\omega[{1,3}]$ & 1.5068 & 2.1016 & 1.9602 & 1.9687 & 2.0108 & 1.4976   \\ \hline\hline
$\omega[{2,2}]$ & 2.6454  & 3.1101 & 3.0039 & 3.0241 & 1.4955 & 2.4976  \\ \hline
$\omega[{2,3}]$ & 2.4965  & 3.0775 & 2.9735 & 3.0066 & 2.4944 & 1.0068  \\ \hline\hline
$\omega[{3,3}]$ & 3.0184 & 2.0464 & 1.9934 & 1.9598 & 1.9845 & 1.9877  \\ \hline
        \end{tabular}}
      % \end{adjustbox}
    \end{table}

%  \begin{table}[t]%[H]
% \caption{Statistics on the error correction of FAID Diversities}
% \centering
% \label{tbl:error_pttrns}
% %\begin{adjustbox}{angle=90}
% %\resizebox{\textwidth}{!}{
% \begin{tabular}{|c|c|c|}         \hline\hline
% Error patterns & FAID Diversity \cite{Declercq_TCOM_2013} & RQNN-aided FAID Diversity \\ \hline 
% ${\cal E}^{(1)}$ & 930 & 0\\ \hline
% ${\cal E}'$ & 507966 & 468844 \\ \hline
%         \end{tabular}%}
%       % \end{adjustbox}
%     \end{table} 
    
\begin{table}[t]%[H]
\caption{Statistics on the error correction of FAID Diversities}
\centering
\label{tbl:error_pttrns}
%\begin{adjustbox}{angle=90}
%\resizebox{\textwidth}{!}{
\begin{tabular}{|c|c|c|}         \hline
Error patterns & FAID Diversity \cite{Declercq_TCOM_2013} & RQNN-aided FAID Diversity \\ \hline 
${\cal E}^{(1)}$ & 930 & 0\\ \hline
${\cal E}'$ & 507966 & 480655 \\ \hline %468844
        \end{tabular}%}
       % \end{adjustbox}
       \vspace{-10pt}
    \end{table} 
    
% \begin{figure}
% \centering
% \includegraphics[width=3.4in]{./figures/FAID_Diversity_FER.eps}
% \caption{FER performance of ${\cal D}_1$, FAID Diversity \cite{Declercq_TCOM_2013}, RQNN-aided FAID Diversity.}
% \label{fig:FER}
% \end{figure}

%% file: Conclusion.tex
\section{Conclusion}
In this paper, we proposed a new approach to automatically design FAID diversity  for LDPC codes over BSC via RQNNs. The RQNN framework models the universal family of FAIDs, thus it can learn arbitrary FAID. %The RQNN framework can construct FAID diversity automatically with large error correction capability and low error floor. 
We applied sub-graph expansion-contraction to sample the most problematic error patterns for constructing the training set, and provided a FER loss function to train the RQNN. We designed the FAID diversity of the Tanner code (155, 64) via the proposed framework. The numerical results showed that RQNN-aided FAID diversity increases the error correction capability and has a low error floor bound.

%% file: ISTC_FAID_NN.bbl
\begin{thebibliography}{10}
\providecommand{\url}[1]{#1}
\csname url@samestyle\endcsname
\providecommand{\newblock}{\relax}
\providecommand{\bibinfo}[2]{#2}
\providecommand{\BIBentrySTDinterwordspacing}{\spaceskip=0pt\relax}
\providecommand{\BIBentryALTinterwordstretchfactor}{4}
\providecommand{\BIBentryALTinterwordspacing}{\spaceskip=\fontdimen2\font plus
\BIBentryALTinterwordstretchfactor\fontdimen3\font minus
  \fontdimen4\font\relax}
\providecommand{\BIBforeignlanguage}[2]{{%
\expandafter\ifx\csname l@#1\endcsname\relax
\typeout{** WARNING: IEEEtran.bst: No hyphenation pattern has been}%
\typeout{** loaded for the language `#1'. Using the pattern for}%
\typeout{** the default language instead.}%
\else
\language=\csname l@#1\endcsname
\fi
#2}}
\providecommand{\BIBdecl}{\relax}
\BIBdecl

\bibitem{hershey2014deep}
J.~R. Hershey, R.~L. Roux, and F.~Weninger, ``Deep unfolding: Model-based
  inspiration of novel deep architectures,'' \emph{arXiv preprint
  arXiv:1409.2574}, 2014.

\bibitem{balatsoukas2019deep}
A.~{Balatsoukas-Stimming} and C.~{Studer}, ``Deep unfolding for communications
  systems: A survey and some new directions,'' in \emph{2019 IEEE International
  Workshop on Signal Processing Systems (SiPS)}, 2019, pp. 266--271.

\bibitem{nachmani2016learning}
E.~Nachmani, Y.~Be'ery, and D.~Burshtein, ``Learning to decode linear codes
  using deep learning,'' in \emph{54th Annual Allerton Conference on
  Communication, Control, and Computing}, Monticello, IL, Oct. 2016, pp.
  341--346.

\bibitem{lugosch2017neural}
L.~Lugosch and W.~J. Gross, ``Neural offset min-sum decoding,'' in \emph{IEEE
  International Symposium on Information Theory (ISIT)}, Aachen, Germany, Jul.
  2017, pp. 1361--1365.

\bibitem{vasic2018learning}
B.~Vasi{\'c}, X.~Xiao, and S.~Lin, ``Learning to decode {LDPC} codes with
  finite-alphabet message passing,'' in \emph{Information Theory and
  Applications Workshop (ITA 2018)}, San Diego, CA, Feb. 2018, pp. 1--10.

\bibitem{XiaoGC2019}
X.~{Xiao}, B.~{Vasi\'c}, R.~{Tandon}, and S.~{Lin}, ``Finite alphabet iterative
  decoding of {LDPC} codes with coarsely quantized neural networks,'' in
  \emph{2019 IEEE Global Communications Conference (GLOBECOM)}, 2019, pp. 1--6.

\bibitem{XiaoTCOM2020}
X.~{Xiao}, B.~{Vasić}, R.~{Tandon}, and S.~{Lin}, ``Designing finite alphabet
  iterative decoders of {LDPC} codes via recurrent quantized neural networks,''
  \emph{IEEE Trans. Commun.,}, vol.~68, no.~7, pp. 3963--3974, 2020.

\bibitem{richardson2003error}
T.~Richardson, ``Error floors of {LDPC} codes,'' in \emph{Proceedings of the
  annual Allerton conference on communication control and computing}, vol.~41,
  no.~3.\hskip 1em plus 0.5em minus 0.4em\relax The University; 1998, 2003, pp.
  1426--1435.

\bibitem{Ivkovic_TIT_2008}
M.~{Ivkovic}, S.~K. {Chilappagari}, and B.~{Vasi\'c}, ``Eliminating trapping
  sets in low-density parity-check codes by using tanner graph covers,''
  \emph{IEEE Trans. Inf. Theory}, vol.~54, no.~8, pp. 3763--3768, 2008.

\bibitem{Declercq_TCOM_2013}
D.~{Declercq}, B.~{Vasi\'c}, S.~K. {Planjery}, and E.~{Li}, ``Finite alphabet
  iterative decoders—part {II}: Towards guaranteed error correction of {LDPC}
  codes via iterative decoder diversity,'' \emph{IEEE Trans. Commun.,},
  vol.~61, no.~10, pp. 4046--4057, 2013.

\bibitem{Planjery2013}
S.~K. {Planjery}, D.~{Declercq}, L.~{Danjean}, and B.~{Vasi\'c}, ``Finite
  alphabet iterative decoders—part {I}: Decoding beyond belief propagation on
  the binary symmetric channel,'' \emph{IEEE Trans. Commun.,}, vol.~61, no.~10,
  pp. 4033--4045, 2013.

\bibitem{Raveendran_TCOM_2020}
N.~{Raveendran}, D.~{Declercq}, and B.~{Vasić}, ``A sub-graph
  expansion-contraction method for error floor computation,'' \emph{IEEE Trans.
  Commun.,}, vol.~68, no.~7, pp. 3984--3995, 2020.

\bibitem{tannercode}
R.~Tanner, D.~Sridhara, A.~Sridharan, T.~Fuja, and J.~Costello, D.J., ``{LDPC}
  block and convolutional codes based on circulant matrices,'' \emph{IEEE
  Trans. Inf. Theory}, vol.~50, no.~12, pp. 2966--2984, Dec. 2004.

\bibitem{karp1972reducibility}
R.~M. Karp, ``Reducibility among combinatorial problems,'' in \emph{Complexity
  of computer computations}.\hskip 1em plus 0.5em minus 0.4em\relax Springer,
  1972, pp. 85--103.

\bibitem{nguyen2018analysis}
T.~T. Nguyen-Ly, V.~Savin, K.~Le, D.~Declercq, F.~Ghaffari, and O.~Boncalo,
  ``Analysis and design of cost-effective, high-throughput {LDPC} decoders,''
  \emph{IEEE Trans. Very Large Scale Integr. (VLSI) Syst}, vol.~26, no.~3, pp.
  508--521, Mar. 2018.

\bibitem{kingma2014adam}
D.~P. Kingma and J.~Ba, ``Adam: A method for stochastic optimization,''
  \emph{arXiv preprint arXiv:1412.6980}, 2014.

\end{thebibliography}
